\RequirePackage{fix-cm}

\documentclass[smallextended]{svjour3}       

\smartqed  
\usepackage{graphicx}

\usepackage[labelfont=bf]{caption}  
\usepackage[labelsep=space]{caption} 

\usepackage[]{graphicx}
\usepackage{textcomp}
\usepackage{booktabs}
\usepackage{amsmath}

\usepackage{multirow}
\usepackage{multicol}
\usepackage{graphicx}
\usepackage{algorithm}
\usepackage{algpseudocode}
\usepackage{enumitem}

\def\Plus{\texttt{+}}
\def\Minus{\texttt{-}}

\usepackage{float}
\usepackage{changepage}
\usepackage{mathtools, cuted}

\usepackage{forest}
\usetikzlibrary{arrows.meta}

\usepackage{tikz}
\newcommand*\circled[1]{\tikz[baseline=(char.base)]{\node[shape=circle,draw,inner
		sep=1pt] (char) {#1};}}

\usepackage{ctable}
\usepackage{subcaption}

\usepackage{tikz-qtree}  

\usepackage{hyperref}
\usepackage{breakurl} 

\newcolumntype{L}[1]{>{\raggedright\let\newline\\\arraybackslash\hspace{0pt}}m{#1}}
\newcolumntype{C}[1]{>{\centering\let\newline\\\arraybackslash\hspace{0pt}}m{#1}}
\newcolumntype{R}[1]{>{\raggedleft\let\newline\\\arraybackslash\hspace{0pt}}m{#1}}

\usepackage[top=1in, left=1in, bottom=1in, right=0.5in]{geometry}

\usepackage[none]{hyphenat}

\usepackage{algpseudocode}  
\usepackage{varwidth}
\usepackage[noadjust]{cite}      

\begin{document}

	\title{Analytical Equations based Prediction Approach for PM2.5 using 
		Artificial Neural Network}

	\author{{Jalpa B Shah$^{1}$}        \and
		Biswajit Mishra$^{2}$
	}

	\institute{Jalpa B Shah \at
		Electronics and Communication Engineering Department, Amrita School of 
		Engineering, Bengaluru 560035, India\\
		\and
		Biswajit Mishra \at
		Dhirubhai Ambani Institute of Information and Communication Technology, 
		Gandhinagar 382007, India \\
		Tel.: +91 079 30510561\\
		\email{biswajit\_mishra@daiict.ac.in}           
		%
	}
	
	\date{Received: date / Accepted: date}

	\maketitle
	
	\begin{abstract}
		
		Particulate matter pollution is one of the deadliest types of air 
		pollution 
		worldwide due to its significant impacts on the global environment and 
		human 
		health. Particulate Matter (PM2.5) is one of the important particulate 
		pollutants to measure the Air Quality Index (AQI). The conventional 
		instruments 
		used by the air quality monitoring stations to monitor PM2.5 are 
		costly, 
		bulkier, time-consuming, and power-hungry. Furthermore, due to limited 
		data 
		availability and non-scalability, these stations cannot provide high 
		spatial 
		and temporal resolution in real-time. To overcome the disadvantages of 
		existing 
		methodology this article presents 
		analytical equations based prediction approach for PM2.5 using an 
		Artificial 
		Neural Network (ANN). Since the derived analytical equations for the 
		prediction 
		can be computed
		using a Wireless Sensor Node (WSN) or low-cost processing tool, it 
		demonstrates 
		the usefulness of the proposed approach. Moreover, the study related to 
		correlation among the PM2.5 and other pollutants is performed to select 
		the 
		appropriate predictors. The large authenticate data set of Central 
		Pollution 
		Control Board (CPCB) online station, India is used for the proposed 
		approach. 
		The RMSE and coefficient of determination (R$^2$) obtained 
		for the proposed prediction approach using eight predictors are 1.7973 
		\textmu 
		g/m$^3$ and 0.9986 
		respectively. While the proposed approach results show RMSE of 7.5372 
		\textmu 
		g/m$^3$ and 
		R$^2$ of 0.9708 using three predictors. Therefore, the results 
		demonstrate that 
		the proposed approach is one of the promising 
		approaches for monitoring PM2.5 without power-hungry gas sensors and 
		bulkier 
		analyzers.

		\keywords{Prediction Model, PM2.5, Correlation, Artificial Neural 
			Network, Air 
			Pollution Monitoring, Machine Learning}
		
	\end{abstract}
	\maketitle

	\section{Introduction}
	Meteorological parameters such as temperature, humidity, light, air 
	velocity 
	and gaseous pollutants such as Carbon Dioxide (CO$_{2}$), Carbon Monoxide 
	(CO), 
	Nitrogen Dioxide (NO$_{2}$), Sulphur Dioxide (SO$_{2}$), Volatile Organic 
	Compounds (VOCs), Suspended Particulate Matter (SPM), and Ozone (O$_{3}$) 
	express the environment quality. The main sources of these pollutants are 
	industrial, residential, transportation, trading, agricultural, and natural 
	activities such as combustion of fuels, wood fire, and forest fire, etc. 
	\cite{airsourcecpcb,airsourceepa}. In India, the Central Pollution Control 
	Board (CPCB) is responsible for providing the ambient Air Quality Index 
	(AQI) 
	through the National Air Quality Monitoring Programme (NAMP). The NAMP 
	\cite{NAMP} network consists of approximately 683 monitoring stations 
	deployed 
	across 300 cities/towns in 29 states and 6 union territories of the 
	country. 
	The objectives of NAMP are to determine the status and trends of ambient 
	air 
	quality and check the prescribed upper and lower limits of the air quality 
	and 
	take corrective measures for the AQI \cite{AQIreport}. The AQI transforms 
	the 
	air quality data into a number, nomenclature, color, and a category;  Good, 
	Satisfactory, Moderately Polluted, Poor, Very Poor and Severe based on 
	their 
	health impacts. The overall AQI \cite{AQI}  is calculated from a minimum 
	three 
	pollutants out of which one should be the Particulate Matter (PM2.5) which 
	can 
	potentially cause serious health problems.
	
	PM2.5 poses a major concern for human health as due to its small size ($<$ 
	2.5\textmu m) they can directly enter into the lungs \cite{WHOreport}. 
	PM2.5 
	comes either from primary sources or from secondary sources.  The primary 
	sources can be vehicles, power plants, wood burning, industrial processes, 
	forest or grass fires, and agricultural burning processes. The secondary 
	sources are precursor emissions such as Sulfur dioxide (SO$_2$), Oxides, 
	Volatile Organic Compounds (VOCs), and Ammonia 
	(NH$_3$) \cite{WHOreport,hodan}. Though different methods and instruments 
	for 
	monitoring PM2.5 exist \cite{amaral2015overview},  only a few are used for 
	real-time measurement and monitoring. These instruments for PM2.5 often 
	lack 
	portability and exhibit a slower response time \cite{budde2014distributed}. 
	Furthermore, the standard procedure to collect samples through samplers and 
	analyzing them offline in specialized laboratories is challenging for 
	real-time 
	monitoring and corrective actions \cite{NAAQsample,pm2.52}. This is because 
	analyzed data is a delayed response of the current data. Hence real-time 
	monitoring of PM2.5 can be useful. In this work, we propose a method to 
	address 
	the issue of the delayed response of data. We propose the PM2.5 prediction 
	model based on analytical equations, which can be ported to a standard 
	Wireless 
	Sensor Node (WSN). We envisage that such a method not only provides 
	benefits 
	for real-time monitoring but also enables an existing WSN to extend its 
	capabilities from monitoring to analyzing.
	
	\section{Related Work}
	
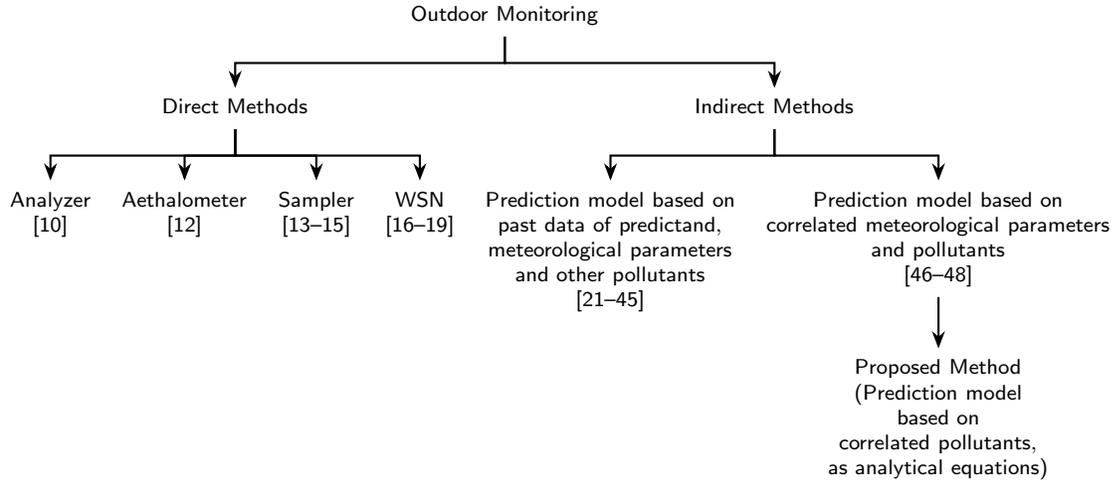
\begin{figure*}
		\centering
		\begin{forest}
			for tree={
				align=center,
				parent anchor=south,
				child anchor=north,
				font=\sffamily,
				edge={thick, -{Stealth[]}},
				l sep+=10pt,
				edge path={
					\noexpand\path [draw, \forestoption{edge}] (!u.parent 
					anchor) -- +(0,-10pt) -| (.child anchor)\forestoption{edge 
						label};
				},
			}
			[Outdoor Monitoring
			[Direct Methods
			[Analyzer \\ \cite{NAAQsample}]
			[Aethalometer \\\cite{Vilcassim2014}]
			[Sampler\\ 
			\cite{gehrig2003characterising,northcross2010estimating,querol2001monitoring}]
			[WSN 
			\\\cite{hall2012portable,mansour2014wireless,kasar2013wsn,liu2011developed}]
			]
			[Indirect Methods
			[Prediction model based on \\ past data of predictand{$,$}\\ 
			meteorological parameters \\ and other pollutants \\ 
			\cite{Ordieres2005,Jef2005,emp6,nn6,lu2006prediction,Sousa2007,gardner1999neural,grivas2006artificial,ann6,
				ann7,ann5, nc8, nc28, A2006, diaz2008hybrid, ann25, clusternn, 
				hybrid5,hybrid7, hybrid6, regularization,cgmmodel7, 
				deeplearning, 
				deeplearning2, deeplearning6}]
			[Prediction model based on \\ correlated meteorological parameters 
			\\ and pollutants \\ \cite{corr27,corr7,everyaware}
			[Proposed Method \\ (Prediction model \\based on \\ correlated 
			pollutants{$,$} \\ as analytical equations)]
			]
			]
			]
			]
		\end{forest}
		
		\caption{Direct and Indirect Methods of Prediction of Pollutants for 
			Outdoor Monitoring} \label{class}
	\end{figure*}

	The classification of both direct and indirect measurements for outdoor 
	environmental monitoring is shown in Fig. \ref{class}. In direct methods, 
	dedicated instruments such as analyzers \cite{NAAQsample}, aethalometer 
	\cite{Vilcassim2014}, samplers  
	\cite{gehrig2003characterising,northcross2010estimating,querol2001monitoring},
	
	and in certain cases Wireless Sensor Nodes (WSNs) 
	\cite{hall2012portable,mansour2014wireless,kasar2013wsn,liu2011developed} 
	are 
	used. Aethalometer and analyzers provide the pollutants' values directly 
	but 
	lack portability and are often expensive. In air pollution monitoring using 
	samplers,  offline analysis is done in specialized laboratories. Whereas in 
	wireless sensor nodes, the measurement of pollutants is done using on-board 
	gas 
	sensors and is often a cost-effective solution compared to other monitoring 
	methods \cite{yi2015survey}. An example of such a system deployed in a New 
	York 
	subway is discussed in \cite{Vilcassim2014}.

	The indirect measurements 
	\cite{Ordieres2005,Jef2005,emp6,nn6,lu2006prediction,Sousa2007,gardner1999neural,grivas2006artificial,ann6,
		ann7,ann5, nc8, nc28, A2006, diaz2008hybrid, ann25, clusternn, hybrid5, 
		hybrid6, regularization,cgmmodel7, deeplearning, deeplearning2, 
		deeplearning6} 
	used prediction approach based on the past data of the predictand, 
	pollutants  
	and/or meteorological parameters. These pollutants and 
	meteorological parameters are correlated with the predictand 
	\cite{hybrid7,corr27,corr7,everyaware}. The PM2.5 (predictand) can be 
	forecasted based on the data of pollutants and meteorological parameters 
	which 
	are correlated with PM2.5. The performance comparison of different 
	prediction 
	techniques of greenhouse gas is discussed in 
	\cite{predictionmethodsreview}. In 
	\cite{Ordieres2005}, a comparison of different topologies of a neural 
	network 
	is presented for a prediction model of PM2.5.
	A neural network-based prediction model for PM10 using previous 
	days data for PM10, cloud cover, boundary layer height, wind direction and 
	day 
	of the week is discussed in \cite{Jef2005}. For the prediction of PM2.5 and 
	O$_3$, the empirical non-linear regression model was designed \cite{emp6} 
	using 
	meteorological parameters and past PM2.5 data.  In \cite{nn6}  
	feed-forward neural network is used for prediction of PM2.5 based on past 
	values of PM10, PM2.5 and some observed and forecasted meteorological 
	parameters. In \cite{lu2006prediction,Sousa2007}, prediction based on  
	ANN using past data of O$_3$ is discussed. A Multilayer Perceptron (MLP) 
	neural 
	network-based prediction model \cite{gardner1999neural} for NO and NO$_2$ 
	pollutants is developed using past data of pollutants. In 
	\cite{grivas2006artificial} 
	prediction model show MLP neural network has better performance than 
	Multiple 
	Linear Regression (MLR) model.
	Results in \cite{ann6} demonstrates the better performance of ANN compared 
	to 
	MLR 
	for prediction of PM2.5 in the agricultural park. Comparison results of MLR 
	and 
	ANN in \cite{ann7}, for prediction of PM2.5 represents the better 
	performance 
	of ANN. Feed Forward 
	Neural Network (FFNN) with Rolling Mechanism (RM) and Accumulated 
	Generating 
	Operation (AGO) of Gray model (RM-GM-FFNN) was developed \cite{ann5} for 
	prediction of PM2.5 and PM10 using past data of PM2.5 and PM10 in addition 
	to 
	meteorological data. Prediction of PM2.5 based on the Back Propagation (BP) 
	neural network was explored \cite{nc8} using satellite-based Aerosol 
	Optical 
	Depth (AOD), meteorological data and past PM2.5 data. The optimized version 
	of 
	the BP network using a genetic algorithm is proposed in \cite{nc28}.

	Few researchers have also explored the hybrid model approach for 
	prediction. A 
	hybrid approach based on the autoregressive and nonlinear model for 
	prediction 
	of NO$_2$ is proposed in \cite{A2006}.  A hybrid approach based on 
	Autoregressive Integrated Moving Average Model (ARIMA) and ANN is discussed 
	in 
	\cite{diaz2008hybrid}. The Comprehensive Forecasting Model (CFM) was 
	developed 
	based on ARIMA, ANN and Exponential Smoothing Method (ESM) \cite{ann25}. 
	Another cluster-based hybrid approach using Neural Network Autoregression 
	(NNAR) and the ARIMA model is discussed in \cite{clusternn} for prediction 
	of 
	PM2.5 using past data of PM2.5. A hybrid-generalized autoregressive 
	conditional 
	heteroskedasticity based prediction approach proposed in \cite{hybrid7} for 
	prediction of PM2.5. In \cite{hybrid5} hybrid model was built for PM2.5 by 
	applying the trajectory-based geographic model and wavelet transformation 
	into 
	the MLP type of neural network. In which meteorological forecasts and 
	pollutants were used as predictors. Comparison of a hybrid model consisting 
	of 
	an Ensemble Empirical Mode Decomposition and General Regression Neural 
	Network 
	(EEMD-GRNN), Adaptive Neuro-Fuzzy Inference System (ANFIS), Principal 
	Component 
	Regression (PCR), and MLR is discussed in \cite{hybrid6} with best results 
	obtained for EEMD-GRNN model. In \cite{regularization} multi-task learning 
	framework is used for the prediction 
	of air pollutants which reduces model parameters with improved performance. 
	Convolutional generalization model implemented \cite{cgmmodel7} for 
	prediction 
	of PM2.5 using meteorological data shows MSE of 15.0 \textmu g/m$^3$. A 
	deep 
	learning-based prediction approach is also implemented for prediction using 
	current and/or previous air pollutants and meteorological data 
	\cite{deeplearning, deeplearning2, deeplearning6}.

	The research work in \cite{corr27} focused on Cuckoo Search-Least Squares 
	Support Vector Machine (CS-LSSVM) based prediction approach for PM2.5 using 
	correlation and principal component analysis. Previous data of PM2.5 was 
	used as one of the predictors in addition to correlated parameters. The 
	correlation analysis of PM2.5 to other meteorological parameters and 
	pollutants using multivariate statistical analysis method and ANN was 
	implemented in \cite{corr7} and prediction results show RMSE of 24.06 
	\textmu g/m$^3$ for ANN-based model.
	Performance comparison of machine learning approaches such as Random 
	Forests 
	(RF), Support Vector Machines (SVMs) and ANN is presented in  
	\cite{everyaware}. Furthermore, a calibration model is developed using ANN 
	for 
	black carbon in which meteorological parameter and other correlated 
	pollutants 
	are used as predictors. The lower RMSE and R$^2$ closeness to 1, also 
	showed 
	the effectiveness of the ANN.

	However, the previously developed prediction model based on past data of 
	predictand does not eliminate the need for dedicated instruments and in 
	almost all cases the proprietary tools are used to measure the predictand. 
	This presents an opportunity for developing a prediction model in the form 
	of analytical equations based on the correlation. The ANN is adopted in our 
	work, due to its superior performance discussed in 
	\cite{lu2006prediction,Sousa2007,gardner1999neural,grivas2006artificial,ann6,ann7,
		everyaware}. Primary results related to the comparison of SVM and ANN 
		are 
	discussed in the results section and demonstrate the effectiveness of our 
	proposed method.\\

	The contribution of the work is as follows:
	
	\begin{enumerate} [leftmargin=0.5cm]
		
		\item The study related to correlation of the pollutants with PM2.5 and 
		additionally the correlation among the pollutants is performed for 
		deciding 
		the predictors.  
		\item Analytical equations are proposed for prediction of PM2.5 using 
		ANN. 
		\item Recalibration of the derived prediction model in terms of 
		coefficients and number of predictors is done to evaluate its 
		performance. 
	\end{enumerate}

	\begin{figure}[t]
		\centering
		\includegraphics[width=3.3in]{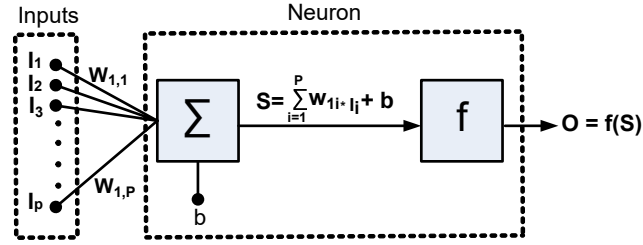}
		\caption{ANN Model of a Neuron}
		\label{neuron}
	\end{figure}
	
	\section{Introduction to ANN Model}

	The proposed prediction model to obtain PM2.5 is based on supervised 
	machine 
	learning. It consists of interconnected computing elements known as neurons 
	with inputs and outputs. As shown in Fig. \ref{neuron}, the model of a 
	neuron 
	has $P$ inputs each with weight $W$. The sum ($S$) of weighted inputs and 
	bias 
	($b$) is fed to the transfer function block ($f$).  The output of each 
	neuron 
	is obtained by subsequently applying the transfer function to the sum of 
	weighted inputs and bias. \\\\\\

	The proposed PM2.5 prediction model using the ANN is derived using the 
	following steps.
	
	\begin{enumerate} 
		[leftmargin=0.9cm,label=\Roman*,labelindent=0cm,itemindent=0pt,labelsep=0.2cm]

		\item Collection of large input data (observations) set for different 
		pollutants. 
		\item Preprocessing of the input data set to remove outliers.
		\item Finding correlation of PM2.5 with other pollutants using 
		preprocessed input data set. 
		\item Based on the correlation results selection of input pollutants 
		(predictors) for developing the prediction model of PM2.5.
		\item Selection of ANN topology for developing the prediction model.
		\item Division of predictors data set into two sets, SET1: 90\% of data 
		set or developing model (training, validation, and testing), SET2: 10\% 
		of data set as unseen data set for further testing.
		\item Developing 100 different ANNs with randomly initialized weights 
		and biases using SET1.
		\item Testing of 100 different ANNs using SET2.
		\item Based on performance indices selection of best ANN for the 
		prediction model.
		\item Deriving analytical equation for prediction using selected ANN.
		\item Prediction of PM2.5 using derived analytical prediction equation.
		
	\end{enumerate}

	\begin{figure*} [b]
		\centering
		\begin{tabular}[c]{cc}
			\begin{subfigure}[c]{0.5\textwidth}
				\includegraphics[width=3.3in,height=2.1in]{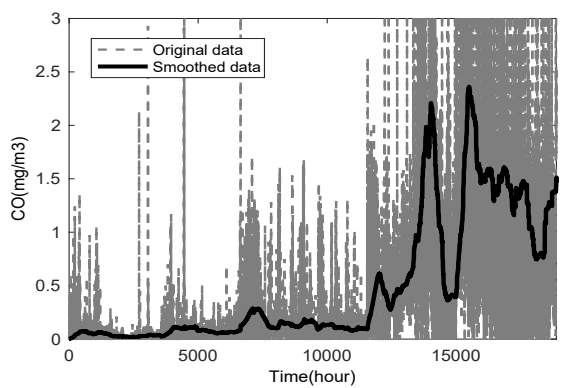}
				\caption{CO Data}\label{co_sma}
			\end{subfigure} &
			
			\begin{subfigure}{0.5\textwidth}
				\includegraphics[width=3.3in,height=2.1in]{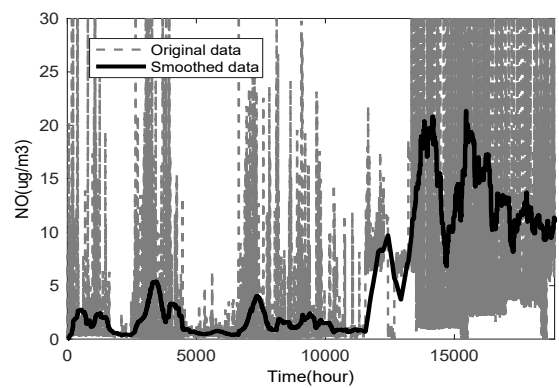}
				\caption{NO Data}\label{no_sma}
			\end{subfigure}\\
			
			\begin{subfigure}[c]{0.5\textwidth}
				\includegraphics[width=3.3in,height=2.1in]{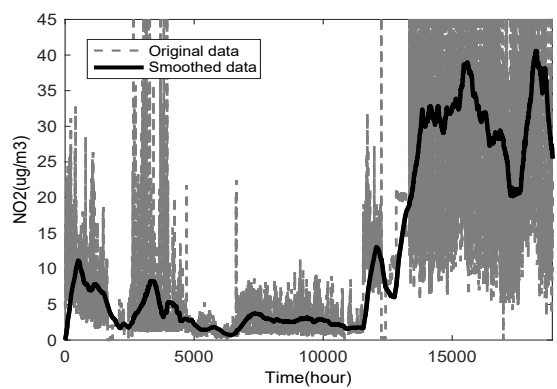}
				\caption{NO$_{2}$ Data}\label{no2_sma}
			\end{subfigure}&
			
			\begin{subfigure}{0.5\textwidth}
				\includegraphics[width=3.3in,height=2.1in]{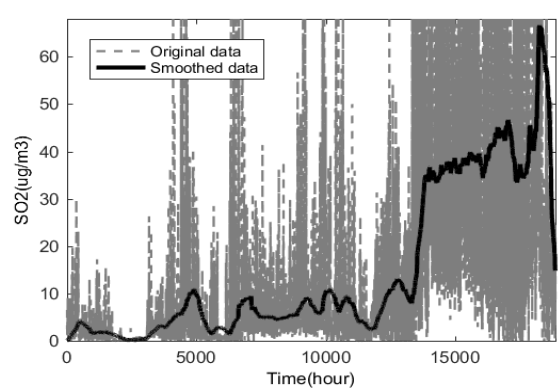}
				\caption{SO$_{2}$ Data}\label{so2_sma}
			\end{subfigure}\\
		\end{tabular}   
		
		\caption{Original and Smoothed Data of CO, NO and NO$_{2}$ and 
			SO$_{2}$} \label{data1}

	\end{figure*}

	\begin{figure*}[p]
		\centering
		
		\begin{tabular}[c]{cc}
			
			\begin{subfigure}[c]{0.5\textwidth}
				\includegraphics[width=3.3in,height=2.1in]{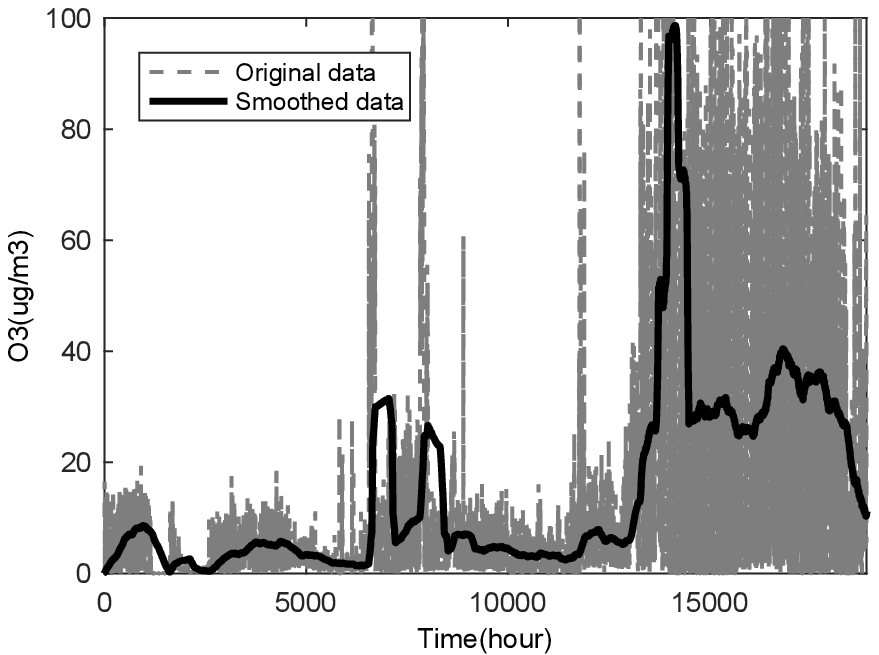}
				\caption{O$_{3}$ Data}\label{no_sma}
			\end{subfigure}&
			\begin{subfigure}{0.5\textwidth}
				\includegraphics[width=3.3in,height=2.1in]{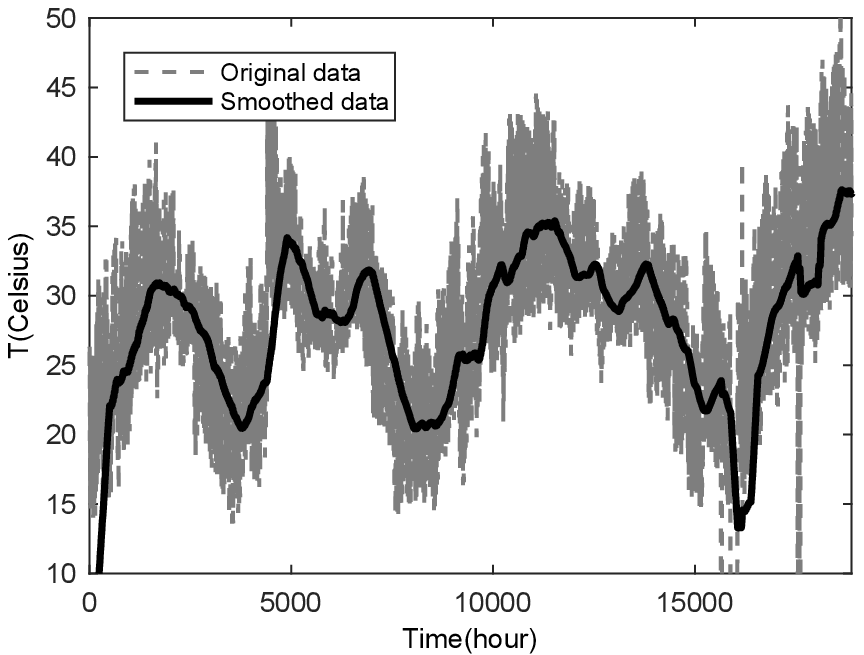}
				\caption{Temperature Data}\label{t_sma}
			\end{subfigure}\\
			
			\begin{subfigure}[c]{0.5\textwidth}
				\includegraphics[width=3.3in,height=2.1in]{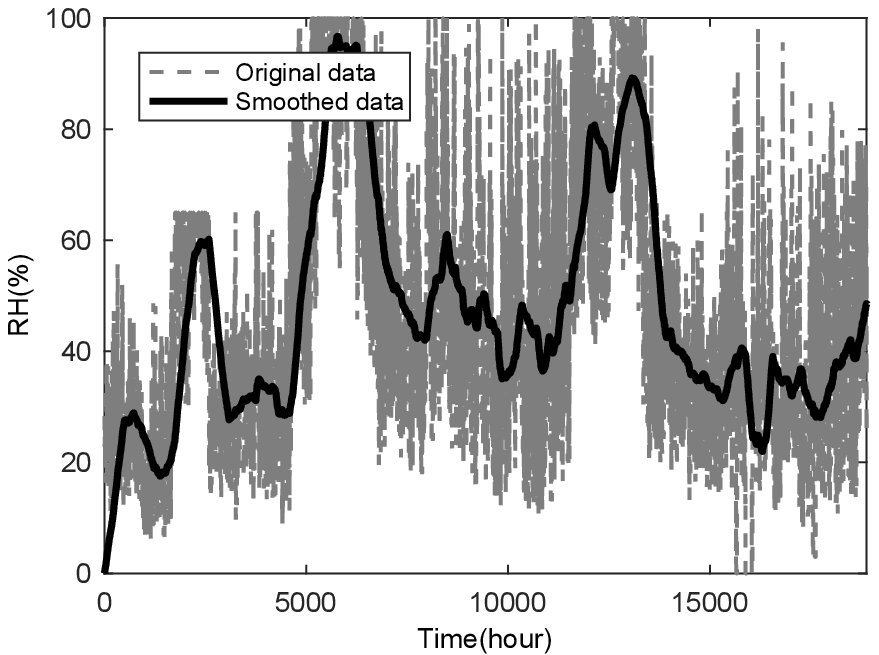}
				\caption{Humidity Data}\label{rh_sma}
			\end{subfigure}&	
			\begin{subfigure}{0.5\textwidth}
				\includegraphics[width=3.3in,height=2.1in]{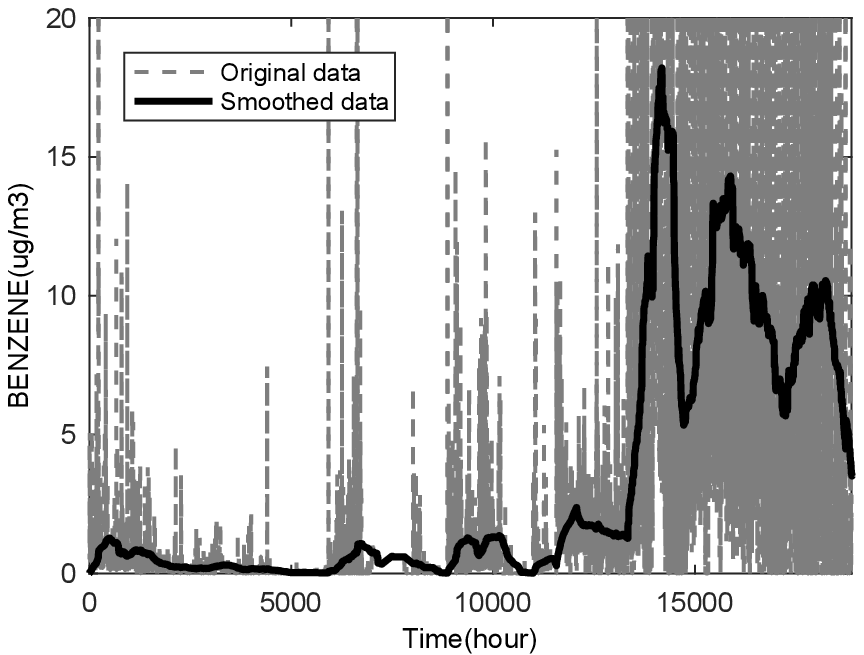}
				\caption{Benzene Data}\label{ben_sma}
			\end{subfigure}\\
		\end{tabular}   
		
		\caption{Original and Smoothed Data of O$_{3}$, Temperature, Humidity 
			and Benzene} \label{data2}

		\begin{tabular}[c]{cc}
			\begin{subfigure}[c]{0.5\textwidth}
				\includegraphics[width=3.3in,height=2.1in]{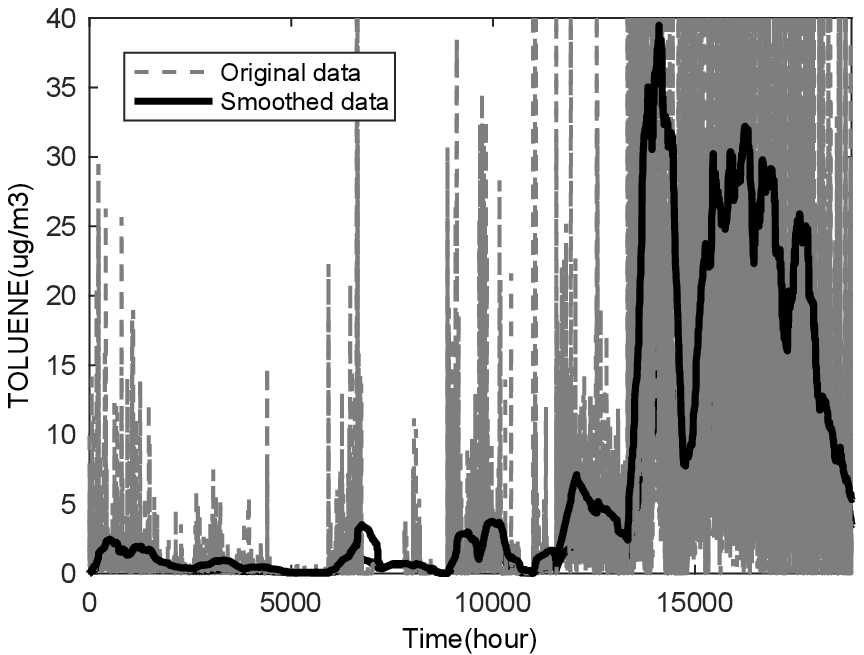}
				\caption{Toluene Data}\label{tol_sma}
			\end{subfigure}&
			\begin{subfigure}{0.5\textwidth}
				\includegraphics[width=3.3in,height=2.1in]{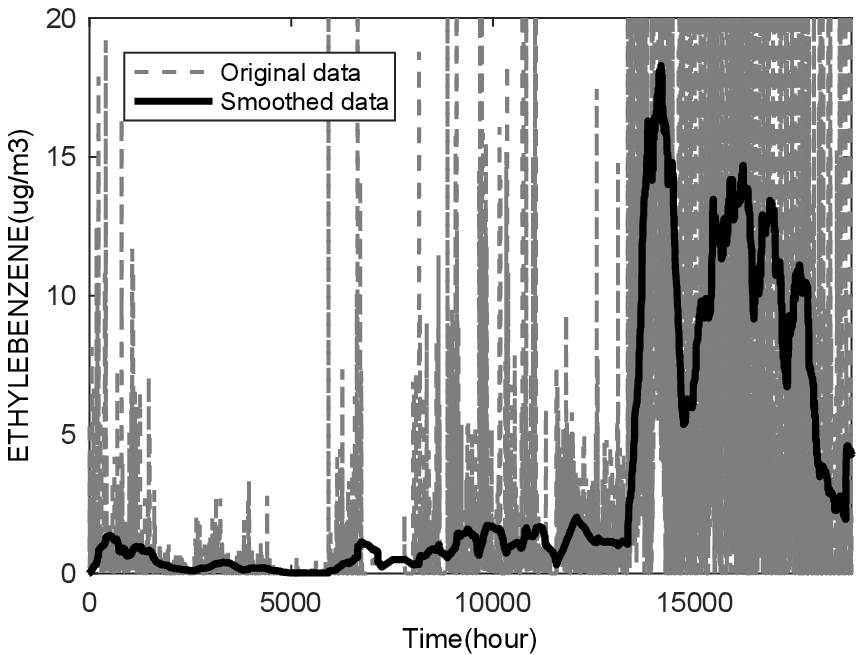}
				\caption{Ethyl Benzene Data}\label{eth_ben_sma}
			\end{subfigure}\\
			\begin{subfigure}{0.5\textwidth}
				\includegraphics[width=3.3in,height=2.1in]{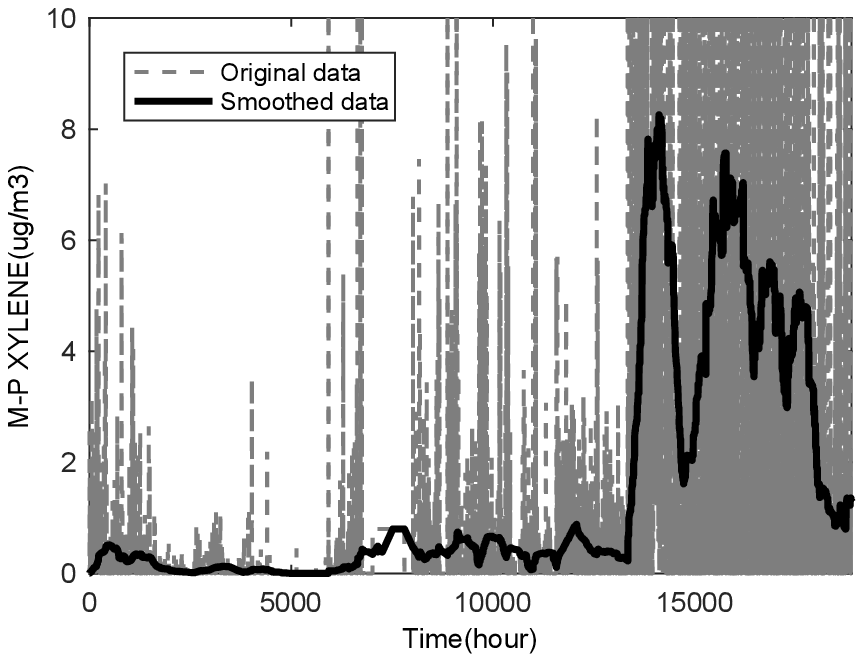}
				\caption{M-P Xylene Data}\label{mp_xyl_sma}
			\end{subfigure}&
			\begin{subfigure}{0.5\textwidth}
				\includegraphics[width=3.3in,height=2.1in]{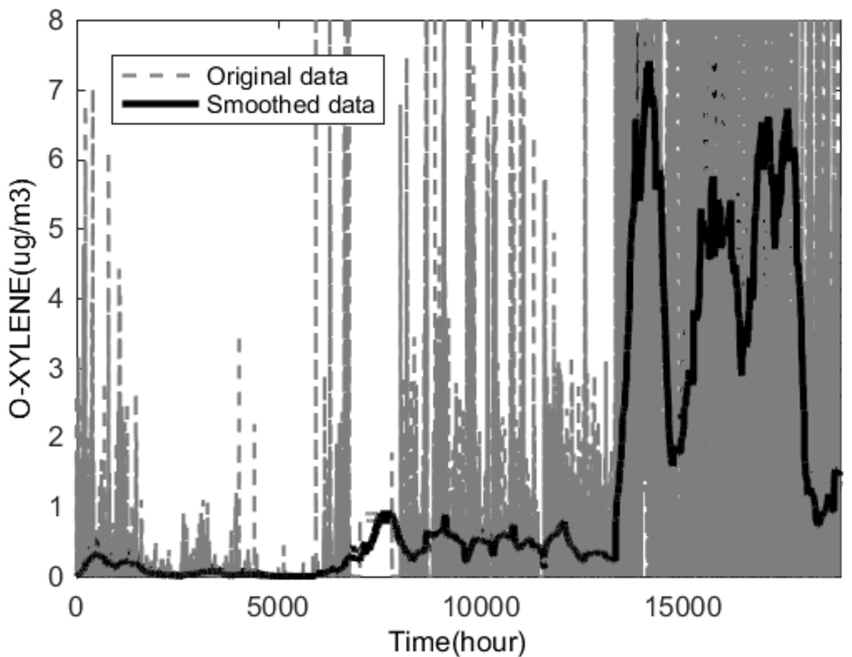}
				\caption{O-Xylene Data}\label{o_xyl_sma}
			\end{subfigure}\\
		\end{tabular}   
		\caption{Original and Smoothed Data of Toluene, Ethyl Benzene, M-P 
			Xylene and O-Xylene} \label{data4}
	\end{figure*}	
	
	\begin{figure}[htb!]
		\centering
		\includegraphics[width=3.3in]{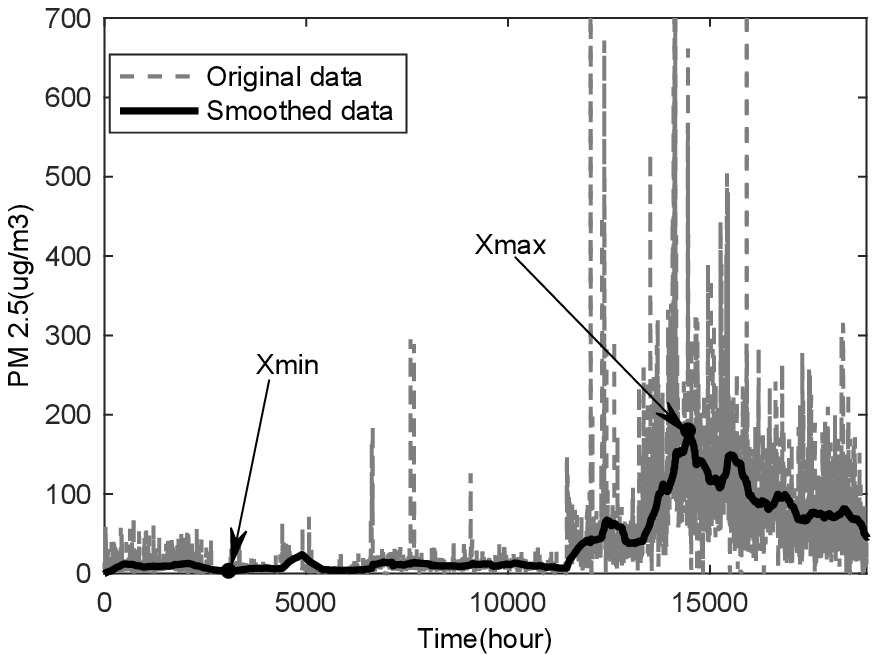}
		\caption{PM2.5 Data}\label{pm_sma}
	\end{figure}

	\section{Observed Data and Correlation}
	
	To select the inputs for the PM2.5 prediction model, it is necessary to 
	obtain 
	the correlation between PM2.5 and pollutants besides the correlation among 
	the 
	pollutants. In the proposed study, Step I is the data collection phase: In 
	this 
	phase, 13 different parameters (pollutants and meteorological parameters) 
	were 
	collected from a CPCB online station, India (N 23$^{\circ}$ 0' 16.6287, E 
	72$^{\circ}$ 35' 48.7816). The data is collected for 41 months, where, the 
	samples of data are taken every hour. Thirteen different parameters 
	monitored 
	from this online station includes pollutants; CO, NO, NO$_2$, SO$_2$, 
	O$_3$, 
	VOC (Benzene, Toluene, Ethyl Benzene, M+P Xylene, O-Xylene), PM2.5 and 
	meteorological parameters; temperature and humidity. Dataset obtained from 
	CPCB 
	online station for all 13 parameters was containing a total of 29,928 
	observations. Due to maintenance, all the parameters data were not 
	available 
	simultaneously for 29,928 observations. After removing maintenance data for 
	each of the parameters, 18,880 observations data set was available 
	simultaneously for all 13 parameters. These 18,880 observations data set 
	was 
	smoothed out to remove the outliers [Step II] and is treated as the golden 
	standard data. The smoothing of the data is done by a moving average 
	filter, 
	which was implemented in MATLAB. The optimum window size for smoothing was 
	found to be 500, resulting in reasonably smoothed data. After smoothing, 
	the 
	first 500 results were removed, as the window size was 500. So, smoothed 
	data 
	of size 18,380 was used for developing the prediction model. Fig. 
	\ref{data1} 
	to Fig. \ref{pm_sma} shows the original and smoothed data for all 13 
	parameters.

	Correlation between any two parameters \textit{x} and \textit{y} is 
	expressed by the correlation coefficient $R$ as per Eq. (\ref{corr1}). 
	\begin{equation}\label{corr1}
	R =\dfrac{ n(\sum xy)-(\sum x)(\sum y)} {\sqrt {(n \sum x ^{2}-(\sum x) 
			^{2}) * (n \sum y ^{2}-(\sum y) ^{2})}}
	\end{equation}
	
	where \textit{n} is the total number of samples. $R$ takes values in the 
	range of [\Minus 1 to \Plus 1]. For a strong positive correlation between 
	\textit{x} and \textit{y}, the value of $R$ will be close to \Plus 1 and 
	vice-versa. For the practical purpose, correlation greater than 0.8 is 
	assumed as being strong and less than 0.5 as weak \cite{corr}.

	\begin{table*}[htb]
		
		\caption{Correlation among the Different Parameters	\label{corr_sma}}
		
		\centering	
		\scalebox{0.88} {	
			\begin{tabular}{ccccccccccccccc}
				\hline
				\textbf{} & \textbf{CO} & \textbf{NO}& \textbf{NO$_2$}&  
				\textbf{SO$_2$}& \textbf{O3} & \textbf{T}& \textbf{RH} & 
				\textbf{Ben}  & \textbf{Tol} & \textbf{Eth} & \textbf{M-xyl} & 
				\textbf{O-xyl} &\textbf{PM2.5}\\
				\hline
				\textbf{CO} & 1 & 0.936 & 0.872 & 0.828 & 0.771 & -0.069 & 
				-0.215 & 0.901 & 0.928 & 0.908 & 0.921 & 0.894 & \textbf{0.830} 
				\\
				\hline
				\textbf{NO} & 0.936 & 	1 & 0.927 & 	0.826 & 0.789 & -0.122 
				& -0.230 & 0.927 & 0.927 & 0.909 & 0.907 & 0.860 & 
				\textbf{0.909} \\
				\hline
				\textbf{NO$_2$} & 0.872 & 	0.927& 	1 &   0.927& 	0.728 & 
				-0.075 & -0.313 & 0.919 & 0.849 & 0.828 & 	0.809 &  0.769 & 
				\textbf{0.909}\\
				\hline
				\textbf{SO$_2$} & 0.828 & 0.826 & 0.927 &  	1 & 0.714 & 0.009 & 
				-0.328 & 0.881 & 0.817 & 0.790 & 0.766 & 0.769& 	
				\textbf{0.845} \\
				\hline
				\textbf{O$_3$} & 0.771 & 0.789 & 0.728 &  0.714 & 1 & -0.066 & 
				-0.238 & 0.844 & 0.849& 0.849 & 0.833 & 0.833& 0.773 \\
				\hline
				\textbf{T} & -0.069 & -0.122 & -0.075&   0.009 & -0.066& 1 & 
				0.294 & -0.126 &-0.158 & -0.191 & -0.217 & -0.127 &   -0.080 	
				\\
				\hline
				\textbf{RH} & -0.215 & 	-0.230 & -0.313 &  -0.328 & -0.238 & 
				0.294 & 1 & -0.318 & -0.297 & -0.323 & -0.315 & 		-0.324 
				& 
				-0.222 \\
				\hline
				\textbf{Benzene} & 0.901 & 0.927 & 0.919&  0.881 & 0.844 & 
				-0.126 & -0.318 & 1 & 	0.956 & 0.943 & 0.934 & 	0.891 & 
				\textbf{0.919} \\
				\hline
				\textbf{Toluene} & 0.928 & 0.927& 0.849 & 0.817& 	0.849 & 
				-0.158 & -0.297 & 0.956 & 1 & 0.989 &  	0.982 &        
				0.955 & \textbf{0.877} \\
				\hline
				\textbf{Ethyl benzene} & 0.908 & 	0.909 & 0.828 &  0.790 & 
				0.849 & -0.191 & -0.323 & 0.943 & 0.989 & 1 & 		0.989 & 
				0.963 & \textbf{0.880} \\
				\hline
				\textbf{M-P	Xylene} & 0.921 & 0.907 & 0.809 &   0.766 & 0.833& 
				-0.217 & -0.315 & 0.934 & 0.982 & 0.989 & 
				1 & 0.964 & \textbf{0.855}\\
				\hline
				\textbf{O-Xylene} & 0.894 & 0.860 & 0.769 &   0.769 & 0.833 & 
				-0.127 & -0.324 & 	0.891 & 0.955 & 0.963 & 
				0.964 & 1 & \textbf{0.818}\\
				\hline
				\textbf{PM2.5} & 0.830 & 0.909 & 0.909 &  0.845 & 0.773 & 
				-0.080 & -0.222 & 	0.919 & 0.877 & 0.880 & 0.855 & 
				0.818 & 1\\
				\hline
			\end{tabular}}
			
		\end{table*}

		Correlation between PM2.5 and other parameters in addition to the 
		correlation 
		among 
		the parameters is evaluated [Step III] based on the 18k\Plus\  data 
		over a 
		period of 41 months. The obtained results are shown in Table 
		\ref{corr_sma}. As 
		can be seen, the highest correlation of PM2.5 is found with NO, NO$_2$ 
		and 
		Benzene and low correlation with temperature and humidity. At typical 
		ambient 
		concentrations, NO is not considered to be hazardous while NO$_2$ can 
		be 
		hazardous \cite{ref1}. Furthermore, NO$_2$ is considered as one of the 
		major 
		pollutants by world health organizations and environmental agencies 
		\cite{ref2, 
			ref3}. Hence, NO$_2$ is considered as one of the predictors for the 
		proposed 
		model and NO is excluded. A strong correlation ($>$0.8) of CO, 
		NO$_{2}$, 
		SO$_{2}$, and VOC (Benzene, Toluene, Ethyl Benzene, M+P Xylene, 
		O-Xylene) with 
		PM2.5 is observed which is useful in selecting the inputs for the PM2.5 
		prediction model. 
		
		The proposed prediction model is based on supervised learning, where, 
		both inputs and target values are provided as a training dataset. So, 
		for the proposed PM2.5 prediction model, as a training dataset, values 
		of input pollutants (CO, NO$_{2}$, SO$_{2}$, and VOC (Benzene, Toluene, 
		Ethyl Benzene, M+P Xylene, O-Xylene)) and target (PM2.5) are taken. It 
		is observed that the training of the ANN will be efficient if each 
		parameter of the training dataset is normalized within the range 
		[-1:1]. Normalization of the parameters is done by finding the value of 
		each parameter within the normalized range using Eq. (\ref{minmaxeq}).

		\begin{equation}\label{minmaxeq}
		x_n =\dfrac{ (Y_{max}-Y_{min}) * (X-X_{min})} { (X_{max}-X_{min})}  
		+Y_{min}
		\end{equation}
		
		where \textit{X} is the value of the parameter while \textit{X$_{max}$} 
		and \textit{X$_{min}$} are the maximum and minimum values of the 
		parameter, respectively. For example, in CO data, \textit{X} is the 
		smoothed value of CO while \textit{X$_{max}$} and \textit{X$_{min}$} 
		are maximum and minimum values of smoothed CO data, respectively. As 
		normalization range is [-1:1], \textit{Y$_{max}$} is 1 and 
		\textit{Y$_{min}$} is -1. The normalized parameters can be converted 
		back into their original form using Eq. (\ref{yminmaxrev}).

		\begin{equation}\label{yminmaxrev}
		X =\dfrac{ (x_n-Y_{min}) * (X_{max}-X_{min})} { (Y_{max}-Y_{min})} 
		+X_{min}
		\end{equation}
		
		where \textit{x$_{n}$} represents normalized data and \textit{X} is the 
		smoothed data. \textit{X$_{max}$} and \textit{X$_{min}$} are the 
		maximum and minimum values of smoothed data, respectively. For example, 
		to convert the normalized data of predictand PM2.5 into original form, 
		\textit{X$_{max}$}  and \textit{X$_{min}$} values of targeted PM2.5, 
		shown in Fig. \ref{pm_sma} are used.

		\begin{figure*}[b]
			\centering
			\includegraphics[width=5in,height=3in]{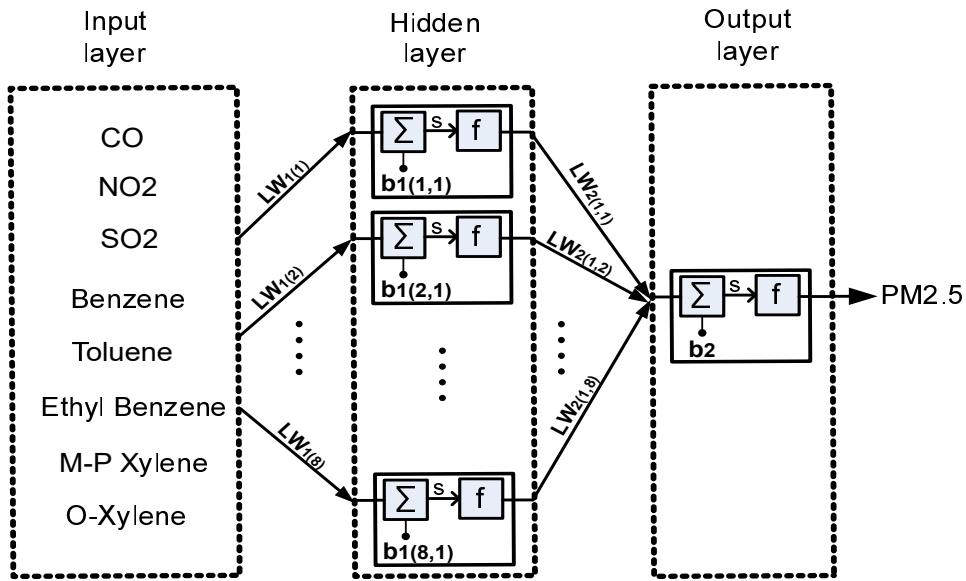}
			\caption{Selected ANN Topology for Prediction}
			\label{nn}
		\end{figure*}

		\section{Prediction model of PM2.5}

		In the proposed prediction model (see Fig. \ref{nn}), inputs are 
		selected based on the correlation results [Step IV]. As can be seen 
		from Table \ref{corr_sma}, a higher correlation ($>$0.8) of CO, 
		NO$_{2}$, SO$_{2}$, and VOC (Benzene, Toluene, Ethyl Benzene, M+P 
		Xylene, O-Xylene) with PM2.5 is observed. So, out of 12 parameters 8 
		parameters; CO, NO$_{2}$, SO$_{2}$, and VOC (Benzene, Toluene, Ethyl 
		Benzene, M+P Xylene, O-Xylene) are selected for developing the PM2.5 
		prediction model. Additionally, as the number of input parameters is 
		reduced from 12 to 8, the computation cost of the proposed prediction 
		model is reduced. The prediction model of PM2.5 shown in Fig. \ref{nn} 
		is based on a feed-forward neural network with a single hidden layer 
		[Step V]. Selected eight pollutants are the input parameters for each 
		neuron of the hidden layer which consists of eight neurons.

		The weights of the hidden layer and output layer can be represented by 
		a matrix of size $S \times P$, where, $S$ is equal to the number of 
		neurons in the layer and $P$ is equal to the number of inputs of the 
		layer. In our case, the matrix size is 8$\times$8, as both the inputs 
		and hidden layer size are 8. The matrix size for the output layer is  
		1$\times$8 as it consists of one neuron and eight inputs coming from 
		the hidden layer. For the proposed predicted model, weights of the 
		hidden layer and output layer are represented by  \textit{LW$_1$} of 
		(of 8$\times$8 size) and  \textit{LW$_2$} of size (of 1$\times$8 size) 
		respectively. The element, \textit{LW$_1${\tiny(1)}} represents the 
		first row of \textit{LW$_1$} matrix, which is formed by the weights of 
		inputs going to the first neuron of the hidden layer. Similarly, 
		\textit{LW$_2${\tiny(1,1)}} represents the first row and first column 
		element of matrix \textit{LW$_2$} and so on. Biases of the particular 
		layer can be represented by a matrix of the size \textit{S}$\times$1, 
		where \textit{S} is equal to the number of neurons in the layer. The 
		bias matrices are \textit{ b$_1$} and \textit{b$_2$} for hidden layer 
		and output layer, respectively. As the number of neurons in the hidden 
		layer is 8 and output layer is 1, the bias matrices, \textit{ b$_1$} is 
		8$\times$1 and \textit{b$_2$} is 1$\times$1, where 
		\textit{b$_1${\tiny(1,1)}} represents the first row and first column 
		element of matrix \textit{b$_1$}.

		The training function \textit{trainlm} based on the Levenberg-Marquardt 
		algorithm is adopted \cite{sharma2014comparison,mathtrain} for 
		training. To train the network for the nonlinear relationship between 
		input and output and to constrain output in positive range standard 
		nonlinear transfer function \textit{logsig} given by Eq. (\ref{logsig}) 
		is used in the hidden layer.

		\begin{equation}\label{logsig}
		logsig(m) =  \dfrac {1} {(1+e^{-m})}
		\end{equation}
		
		where \textit{m} is the input to the transfer function. In 
		\textit{logsig} transfer function, the output will be in the range of 
		[0:1] for the entire range of inputs. For nonlinear regression or 
		prediction, \textit{purelin} is an effective transfer function for the 
		output layer \cite{ref4}. Hence in the proposed model, \textit{purelin} 
		is used as a transfer function in the output layer. In 
		\textit{purelin}, the output will be equal to the input.
		
		The total available data of 18k\Plus\ is divided into two sets the 
		first set (SET1) includes 90\% of data and the second set (SET2) 
		includes 10\% of data [Step VI]. The division into two sets is done 
		randomly so that all types of data can be included in two sets. The 
		SET1 (90\% data) is used for designing the neural network and it is 
		divided in a standard manner widely used by researchers, 70\% for 
		training, 15\% for validating and 15\% for testing. The SET2 (10\% 
		data) is kept as unseen data for further testing and comparing the 
		performance of neural networks.

		\begin{algorithm}
			
			\begin{algorithmic}[1]
				
				\State Load the data SET1 for developing the network and SET2 
				to test the Network for unseen data
				\For {i = 1 to 100}
				\State \begin{varwidth}[t]{\linewidth}
					Divide SET1 randomly for training(70\%), validating(15\%)\\ 
					\hskip\algorithmicindent and testing (15\%)\
				\end{varwidth}
				\State Initialize the weights and biases 
				\State Train the network
				\State Validate the network
				\State Test the network
				\State Evaluate training, validation and testing performance
				\State Save the performance results
				\State Save Network and its weights and biases
				\State  \begin{varwidth}[t]{\linewidth}Test the Network 
					performance for unseen data using SET2 \end{varwidth}
				\State Save the performance results
				\EndFor
				
			\end{algorithmic}
			\caption{Pseudo Code for Developing and Testing 100 different 
				ANNs}\label{code}
			\label{code}
		\end{algorithm}

		For selecting the best ANN for prediction model, 100 different ANNs are 
		developed and tested as per the pseudo-code in Algorithm \ref{code}. 
		This algorithm is repeated 100 times to get the performance of 100 
		different ANNs for comparing training and testing results with good 
		generalization [Step VII, VIII]. The performance of the ANN is 
		evaluated [Step IX] based on, RMSE and R$^2$, where, RMSE is the root 
		mean square of the errors, i.e, the difference between the target value 
		and the predicted value and is given in Eq. (\ref{rmse}). 
		
		\vspace{-1em}
		\begin{equation}\label{rmse}
		RMSE =\sqrt {\dfrac{\sum_{i=1}^{n} (A_i -P_i)^{2}} {n}}
		\end{equation}
		
		where \textit{A$_{i}$} represents the actual value, \textit{P$_{i}$} 
		represents the predicted value and \textit{n} is the total number of 
		samples. R$^2$ is the coefficient of determination and it is the square 
		of the correlation R (Eq. (\ref{corr1})). The closeness between target 
		values and predicted outputs of ANN is represented by R$^2$. The value 
		of R$^2$ equal to 1 represents targets and predicted outputs are very 
		close to each other.
		\subsubsection{Model Equations}
		
		The hidden layer and output layer weights for the proposed model are 
		shown in Eq. (\ref{lw1}) and Eq. (\ref{lw2}) respectively. The biases 
		of the model for hidden layer and output layer are shown in Eq. 
		(\ref{b1}) and Eq. (\ref{b2}) respectively. Matrix \textit{B$_{1}$} is 
		formed by repeating the hidden layer bias matrix \textit{b$_{1}$}, N 
		times, where N is equal to the size of test dataset (1838 in our case). 
		This operation shown in Eq. (\ref{B1}), is performed  to make bias 
		matrix size \textit{B$_{1}$} equal to the size of product term matrix 
		\textit{{$LW_1$}*{$x_n$}}, so that addition operation (\textit{B$_{1}$} 
		+ \textit{{$LW_1$}*{$x_n$}}), shown in Eq. (\ref{pmmodel}) can be 
		performed.

		\begin{equation}\label{lw1}
		\hspace*{-0.85cm} 
		LW_1  =
		\left[\begin {matrix}
		-25.978 &	-31.063 &	17.505 &38.446	 & -37.353 & 	97.708 & 
		-92.951 &	65.964 \\
		3.295 &	6.460 &	0.203 &	-1.981 &	10.078 &	-10.281 &	3.352 &	
		-3.782\\
		-2.661 & 2.503 &	-1.075 &-1.011 &-0.194&	1.702 &	-2.768 &	1.690\\
		-6.069 &	6.142 &	-2.212 &-3.593 &	1.800 &	-3.002 &	0.222 &	
		2.803\\
		-9.788 &	-7.748 &	-4.424&	27.227 &	-4.406 &	0.016 &	-6.279 
		&	8.460\\
		3.821&	-6.233 &	1.243	& -1.449& 	5.600 &	1.846 &	1.801 &	-5.862\\
		3.149&	6.319 &	0.368 &	-0.825 &	10.746 &	-11.220 &	2.650 & 	
		-3.428\\
		-2.127 &	1.991&	-1.116 &	-0.032 &	0.038 &	4.094 &	-4.882 &	
		1.330\\
		\end{matrix}\right]
		\end{equation}

		\begin{equation}\label{lw2}
		\hspace*{-0.9cm} 
		LW_2  =
		\left[\begin {matrix}
		0.158 &	-17.591 &	-6.481 &	1.614 &	-0.369 &	-0.817 &	17.345 
		&	4.950\\
		\end{matrix}\right]
		\end{equation}

		\begin{equation}\label{b1}
		b_1  =
		\left[\begin {matrix}
		34.474\\
		4.001\\
		-1.015\\
		-2.615\\
		5.700\\
		-0.148\\
		4.383\\
		-1.959\\
		\end{matrix}\right]
		\end{equation}

		\begin{equation}\label{b2}
		b_2  =
		\left[\begin {matrix}
		1.593\\
		\end{matrix}\right]
		\end{equation}

		\begin{equation}\label{B1}
		B_1  =
		\left[\begin {matrix}
		b_1 & b_1 & b_1 & b_1 & .& . & . &.& .&N
		\end{matrix}\right]
		\end{equation}

		\vspace{10mm}
		
		The prediction model for deriving PM2.5 [Step X] is given by Eq. 
		(\ref{pmmodel}).
		\begin{equation}\label{pmmodel}
		PM2.5_n = b_2 + LW_2 * logsig( B_1 + LW_1 * x_n )
		\end{equation}
		
		where \textit{PM2.5$_{n}$} is normalized output.
		The input matrix \textit{x$_{n}$} is formed using normalized values of 
		CO, NO$_{2}$, SO$_{2}$, and VOC and the size of \textit{x$_{n}$} is 
		$P\times N$ where $P$=8 is the number of inputs and $N$=1838 is the 
		number of values for each input. Eq. (\ref{pmmodel}) shows the 
		usefulness of the approach to obtain the value of PM2.5 based on CO, 
		NO$_{2}$, SO$_{2}$, and VOC values only. Therefore, low-cost sensors 
		can be deployed with the derived model that offers the opportunity to 
		derive PM2.5 through signal processing algorithms. Eq. (\ref{pmmodel1}) 
		shows the conversion to the original value from the normalized one 
		using Eq. (\ref{yminmaxrev}), in which \textit{X$_{max}$} and 
		\textit{X$_{min}$} are taken as per Fig. \ref{pm_sma}.

		\begin{equation}\label{pmmodel1}
		\hspace*{-0.2cm} 
		PM2.5 =\dfrac{ (PM2.5_n+1) * (180.052-2.228)} {2} +2.228
		\end{equation}

\begin{figure*}[htb!]
				\centering
				\includegraphics[width=5in,height=2.5in]{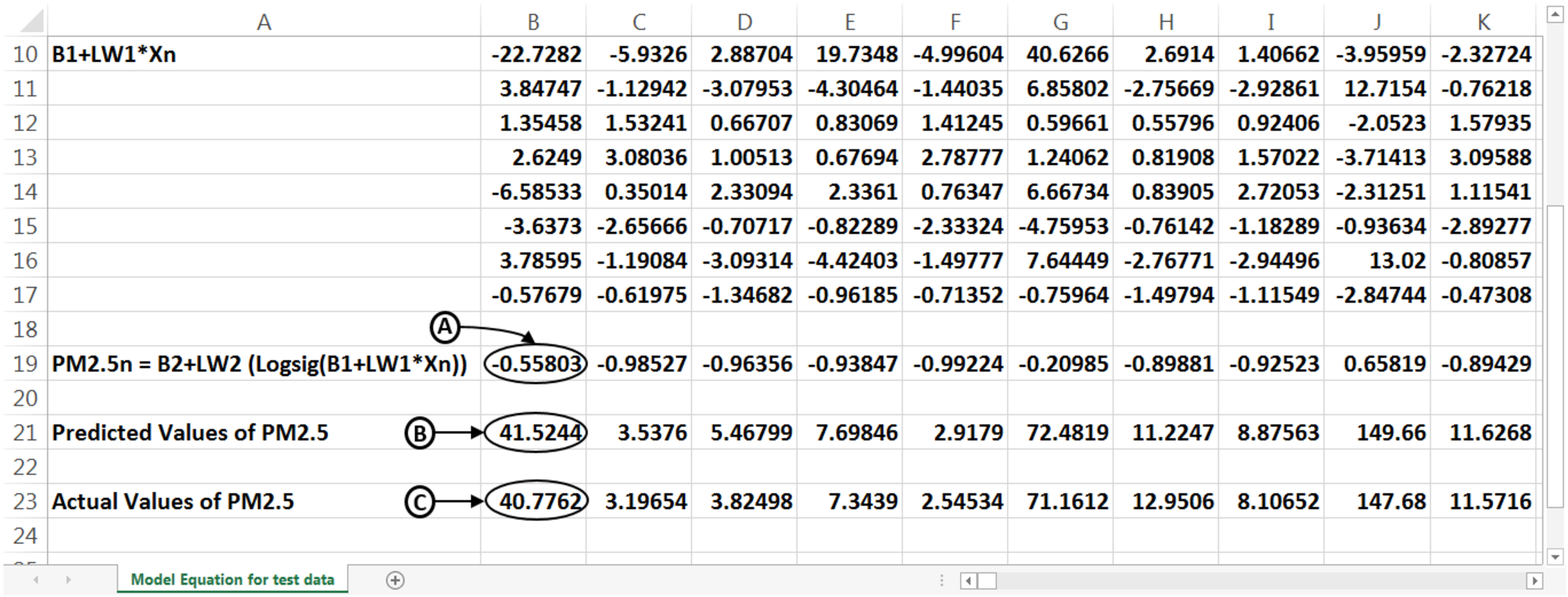}
				\caption{Excel sheet used for processing prediction model 
					equation}
				\label{excel_model}
			\end{figure*}

		The proposed approach based on the above analytical equations 
		eliminates the 
		need for proprietary tools. The analytical equation for the prediction 
		can be 
		computed using any low-cost processing tool (e.g. excel sheet etc.). A 
		screenshot of an example [Step XI] is shown in Fig. \ref{excel_model}. 
		In this 
		example, an input matrix \textit{x} of size 8$\times$1838 is taken 
		which 
		represents the testing data set. Eq. (\ref{pmmodel}) is computed as 
		follows. 
		
		\begin{enumerate} 
			[leftmargin=0.9cm,labelindent=0cm,itemindent=0pt,labelsep=0.2cm] 
			
			\item [a.] First, \textit{$x_n$} is obtained by computing Eq. 
			(\ref{minmaxeq})
			\item [b.] Multiplying the two matrices of \textit{$LW_1$} (Eq. 
			(\ref{lw1})) and \textit{$x_n$}, product term 
			\textit{{$LW_1$}*{$x_n$}} is obtained.
			\item [c.] Matrix $B_1$, formed using Eq. (\ref{B1}) is added in 
			the product term \textit{{$LW_1$}*{$x_n$}}.
			\item [d.] Hidden layer activation function (Eq. (\ref{logsig})) is 
			applied to find the output of the hidden layer. 
			\item [e.] The output of the hidden layer is multiplied by 
			\textit{$LW_2$} (Eq. (\ref{lw2})).
			\item [f.] Then matrix \textit{$b_2$} from Eq. (\ref{b2}), is 
			added. 
			\end {enumerate}
			
			Finally, PM2.5$_n$ (marked as \circled{A} in Fig. 
			\ref{excel_model}) is 
			obtained by computing Eq. (\ref{pmmodel}), which is the normalized 
			value of 
			PM2.5. PM2.5 in the original unit is obtained by processing Eq. 
			(\ref{pmmodel1}) (marked as \circled{B} in Fig. 
			\ref{excel_model}).  
			Predicted values of PM2.5 obtained from Eq. (\ref{pmmodel1}) 
			(marked as 
			\circled{B} in Fig. \ref{excel_model}) are close to the actual 
			values of 
			PM2.5 (marked as \circled{C} in Fig. \ref{excel_model}). 
			Interestingly 
			these computations can be ported to a wireless sensor node having 
			basic 
			memory along with computational capability and the algorithms can 
			still 
			perform 
			reliably.

			\begin{figure*}[htb!]
				\centering
				\includegraphics[width=5in]{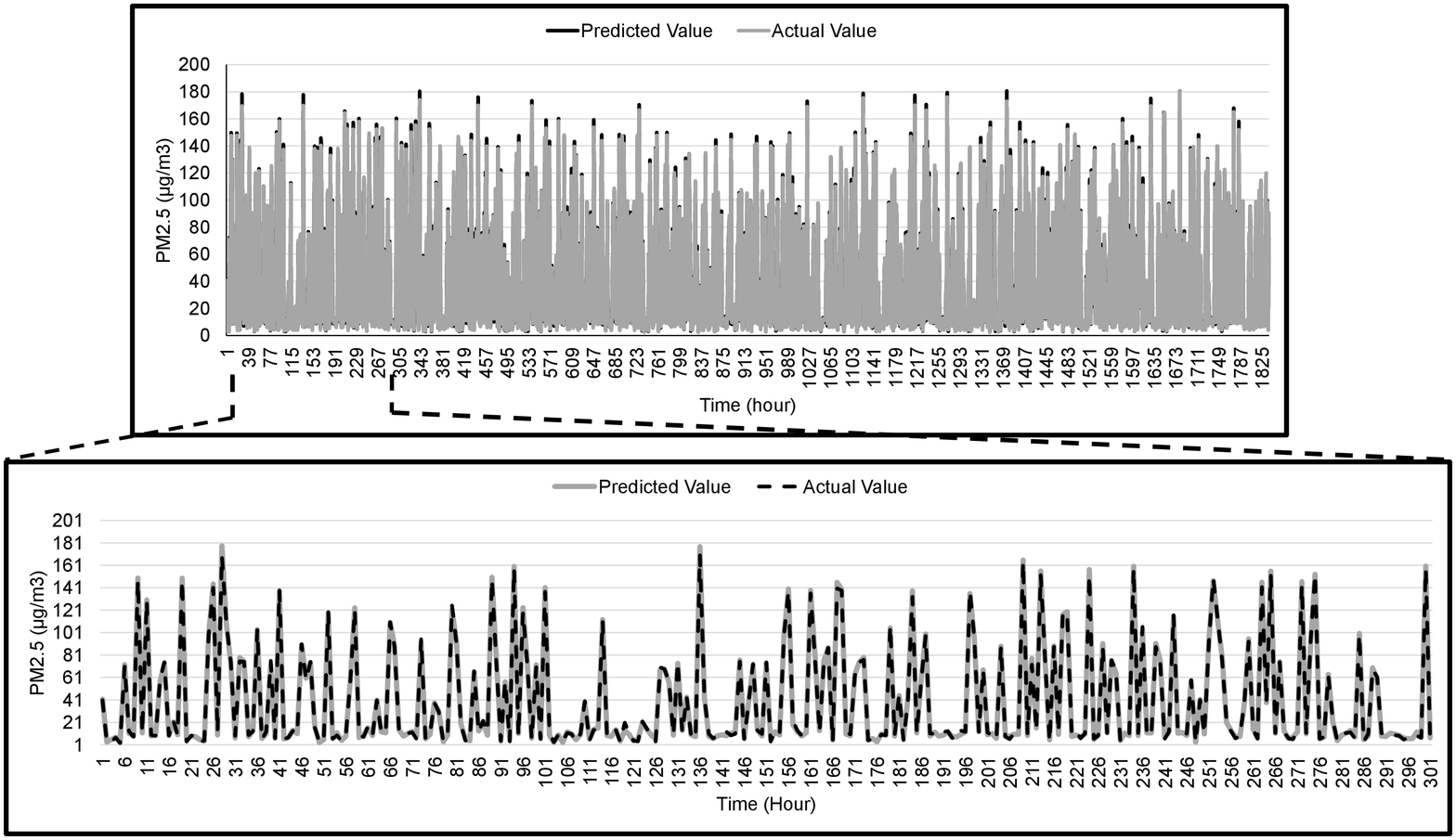}
				\caption{Good match of the results obtained from the analytical 
					equation of model with actual values}
				\label{results}
			\end{figure*}

			The deployment of the developed prediction model needs to consider 
			the following limitations based on the location, available 
			monitoring stations and available monitoring parameters as 
			predictors.
			
			\renewcommand{\labelitemi}{$\bullet$}
			\begin{itemize}
				[labelindent=0pt,itemindent=0pt,leftmargin=0.8cm,topsep=2pt, 
				labelsep=0.2cm]

				\item Air pollution varies from one location to another based 
				on the few parameters like human activity, traffic condition, 
				the structure of urban area and weather conditions.  Based on 
				the location, predictors and predictand values as well as their 
				maximum and the minimum limit will change.  So, the application 
				of the presented prediction model for another location needs 
				model training, validation, and testing again which provides 
				new coefficients in terms of weights and biases for accurate 
				prediction with lower RMSE.
				\item Application of prediction model to another location 
				requires a large set of authentic data for training which is 
				sometimes difficult due to the limitation on the number of 
				online environment monitoring stations.  Due to a few 
				monitoring parameters, and delay in the availability of data; 
				offline stations are less preferable than online stations. 
				\item Online stations of CPCB monitors pollutants that are 
				greater than the offline stations.  Concentration data for a 
				large set of pollutants is the basic requirement for a 
				correlation study or developing a prediction model. Due to the 
				limitation on the online stations available in the city, data 
				were used from only one online station for training, 
				validation, and testing in the proposed study.  This can be 
				expanded in the future by taking data from multiple online 
				stations of different cities.
				
				\item The derived prediction model based on ANN generally shows 
				poor performance for predicting the sudden large change in 
				predictors.  As, sometimes it is difficult to discriminate 
				between the outliers and sudden change in the value, applying a 
				smoothing algorithm to remove outliers will also remove the 
				sudden large change in the value of the predictor.  This 
				limitation can be targeted through different data processing 
				algorithm and is left as part of future work

			\end{itemize}

			\subsubsection{Model Results}
			
			In comparison to Support Vector Machine (SVM), the ANN exhibits 
			better performance in terms of RMSE. The RMSE of 2.862 and 2.823 is 
			obtained during training and testing respectively for the SVM model 
			(shown in Table \ref{svm}).  While for ANN, the RMSE of 1.5971 and 
			1.5121 is obtained (shown in Table \ref{per2}) during training and 
			testing for unseen data respectively.

			\begin{table}[b!]
				
				\caption{Performance of SVM \label{svm}}
				
				\centering
				\begin{tabular}{C{5cm}C{2cm}}
					\hline
					\textbf{Performance of} & \textbf{RMSE}\\
					\hline
					\hline
					\textbf{Training}	& 2.862 	  \\
					\hline
					\textbf{Testing for unseen data}	& 2.823	  \\
					\hline
				\end{tabular}
			\end{table}

			\begin{table}[htb!]
				
				\caption{Performance of Prediction Model \label{per2}}

				\centering
				\begin{tabular}{C{4cm}C{1.5cm}C{1.5cm}}
					\hline
					\textbf{Performance of} & \textbf{RMSE} & \textbf{R$^2$}\\
					\hline
					\hline
					\textbf{Training}	& 1.5971 &	0.9987 \\
					\hline
					\textbf{Validation}	& 1.6347&	0.9986 \\
					\hline
					\textbf{Testing}	& 1.5843&	0.9988 \\
					\hline
					\textbf{Testing for unseen data}	& 1.5121	 & 0.9988 \\
					\hline
				\end{tabular}
			\end{table}
			
			Table \ref{per2} shows the best performance of the prediction model 
			obtained from 100 iterations. The value of R$^2$ demonstrates the 
			closeness of the predicted values with the target values or actual 
			values. The actual values of PM2.5 are compared (see Fig. 
			\ref{results}) with predicted results obtained by Eq. 
			(\ref{pmmodel1}). It is found that the predicted results are in 
			close accordance with actual values, which confirms the 
			effectiveness of the proposed approach.

			\begin{table}[htb!]
				
				\caption{Summary of previously developed prediction 
					models\label{comparision}}
				
				\centering		
				\begin{tabular}{cccc}
					\hline
					\textbf{Reference} & \textbf{Predictand} & \textbf{RMSE} & 
					\textbf{R$^2$}\\
					\hline
					\hline
					\cite{lu2006prediction} & O$_3$ (ppb) & 0.30 & 0.69 \\ 
					\hline
					\cite{Sousa2007}& O$_3$ (\textmu g/m$^3$) & 21.78 & 0.73 \\
					\hline
					\cite{gardner1999neural}& NO$_2$ (ppb) & 7.3 &  0.91 \\
					\hline
					\cite{A2006} & NO$_2$ (\textmu g/m$^3$) & 13.93 &  0.93 \\
					\hline
					\cite{grivas2006artificial} & PM 10 (\textmu g/m$^3$) & 
					12.16 &  0.83 \\
					\hline
					\cite{diaz2008hybrid}& PM 10 (\textmu g/m$^3$) & 11.656 & 
					0.983 \\
					\hline
					\cite{nc8}& PM2.5 (\textmu g/m$^3$) & 41.97 & - \\
					\hline
					\cite{ann25}& PM2.5 (\textmu g/m$^3$) & 12.8903  & - \\
					\hline
					\cite{ann5}& PM10 (\textmu g/m$^3$) & 18.4 & 0.895 \\
					\hline
					\cite{ann5}& PM2.5 (\textmu g/m$^3$) & 12.7  & 0.954 \\
					\hline
					\cite{ann6}& PM2.5 (\textmu g/m$^3$) & 6.77 & 0.99 \\
					\hline
					\cite{hybrid6}& PM2.5 (\textmu g/m$^3$) & 5.0324 & 0.79 \\
					\hline
					\cite{corr27}& PM2.5 (\textmu g/m$^3$) & 14.47 & - \\
					\hline
					\cite{corr7}& PM2.5 (\textmu g/m$^3$) & 24.06 & - \\
					\hline
					\cite{everyaware} & BC (ng/m$^3$) & 1480.746 & 0.586\\
					\hline
					This Work   & PM2.5 (\textmu g/m$^3$) & 1.7973 & 0.9986 \\
					\hline
				\end{tabular}
			\end{table}

			Previously developed prediction models, depend on past data, are 
			often time-consuming and use dedicated instruments.  In the 
			prediction model, we have eliminated the above requirements and 
			comparison with some researches is shown in Table \ref{comparision}.
			As can be seen, the proposed approach has better RMSE and R$^2$  
			compared to existing methods. The RMSE of 1.7973 \textmu g/m$^3$ 
			and R$^2$ of 0.9986 is obtained for test dataset of size 1838. We 
			extended the model to accommodate a reduced test dataset of size 
			10, which shows RMSE of 146.10 \textmu g/m$^3$ and R$^2$ of 0.9467. 
			The proposed model in the form of the analytical equation thus 
			helps in predicting PM2.5 using low-cost processing tools or 
			existing WSN.

			The proposed prediction model is recalibrated in terms of the 
			number of 
			predictors, weights, and biases to show the effectiveness of the 
			proposed 
			approach. Instead of eight predictors, three predictors, CO, 
			NO$_{2}$, and 
			Benzene (VOC component) are taken considering the availability of 
			low-cost 
			sensors \cite{mics6814} which includes this type of multiple 
			sensing 
			parameters. The proposed approach can work for any other three 
			sensing 
			parameters after recalibrating model. In recalibration, CPCB 
			smoothed data 
			for CO, NO$_{2}$, and Benzene (VOC component) are used as it is 
			considered 
			as golden standard data. For recalibration ANN shown in Fig. 
			\ref{nn} is 
			used with the same training and testing dataset (of three 
			parameters), but 
			the size of the input layer and the hidden layer is reduced to 
			three. 
			Extracted weights and biases are represented in Eq. (\ref{lw11}) to 
			Eq. 
			(\ref{b22}). Performance results are shown in Table \ref{per3}. 
			Results 
			show, during testing RMSE is 7.5372 \textmu g/m$^3$ and R$^{2}$ is 
			0.9708.
			
			\begin{equation}\label{lw11}
			\hspace*{-0.85cm} 
			LW_1  =
			\left[\begin {matrix}
			26.281 &	3.456 &	-12.391\\
			17.898 &	-0.863&	11.305\\
			-0.996&	-0.502&	-1.205\\
			\end{matrix}\right]
			\end{equation}
			
			\begin{equation}\label{lw22}
			\hspace*{-0.9cm} 
			LW_2  =
			\left[\begin {matrix}
			-1.008 &	1.379 & -1.665 \\
			\end{matrix}\right]
			\end{equation}

			\begin{equation}\label{b11}
			b_1  =
			\left[\begin {matrix}
			9.934\\
			21.939\\
			1.101\\
			\end{matrix}\right]
			\end{equation}

			\begin{equation}\label{b22}
			b_2  =
			\left[\begin {matrix}
			0.689\\
			\end{matrix}\right]
			\end{equation}

			\begin{table}[htb!]
				
				\caption{Performance of Prediction Model for Three Predictors 
					\label{per3}}
				
				\centering
				
				\begin{tabular}{C{4cm}C{1.5cm}C{1.5cm}}
					\hline
					\textbf{Performance of} & \textbf{RMSE} & \textbf{R$^2$}\\
					\hline
					\hline
					\textbf{Training}	& 7.9297 	&	0.9678 \\
					\hline
					\textbf{Validation}	& 8.0954 	&	0.9679 \\
					\hline
					\textbf{Testing}	&8.0049 	&	0.9672 \\
					\hline
					\textbf{Testing for unseen data}	& 7.5372	&	0.9708 
					\\
					\hline
				\end{tabular}
			\end{table}

			\section{Conclusion and Future Work}
			
			In this work, we have studied the correlation of PM2.5 with other 
			pollutants and correlation among the pollutants, based on the 
			standardized 
			CPCB data. The prediction model of PM2.5 is based on the 
			correlation and 
			shown to be useful for online as well as offline measurements. The 
			proposed 
			model is in the form of analytical equations that enables the use 
			of any 
			low-cost processing tool and eliminates the need for a proprietary 
			tool for 
			predicting PM2.5 values. Results obtained using this method with 
			eight 
			predictors ( NO$_{2}$, SO$_{2}$, and VOC (Benzene, Toluene, Ethyl 
			Benzene, 
			M+P Xylene, O-Xylene)) shows RMSE of 1.7973\textmu g/m$^3$ and 
			R$^2$  
			0.9986 for the test dataset. To show the effectiveness of the 
			proposed 
			approach, the derived prediction model was recalibrated with three 
			predictors (CO, NO$_{2}$, and Benzene (VOC component) due to the 
			possibility of sensing all three parameters by one or two low-cost 
			sensors. 
			The proposed approach can work for any other three sensing 
			parameters too 
			after recalibration. Testing results show  RMSE of  7.5372 \textmu 
			g/m$^3$ 
			and R$^2$ of 0.9708. The obtained results prove the effectiveness 
			of the 
			proposed approach. The obtained results can be improved in the 
			future by 
			recalibrating prediction model based on the data available from 
			multiple 
			stations located at the place of deployment. In comparison to 
			existing 
			methods, the proposed approach facilitates an efficient method that 
			reduces 
			overall computation cost. Furthermore, this model can be 
			implemented on the 
			wireless sensor node for automated measurement of PM2.5. 
			
			\section{Acknowledgement}
			The authors thank the Central Pollution Control Board (CPCB) and 
			their staff for their support during this research work.
			
			\section*{Compliance with ethical standards}
			\textbf{Conflict of interest:} The authors declare that they have 
			no conflict of
			interest.

		\end{document}